\documentstyle[epsfig,amsfonts]{article}
\begin{document}

\title{A geometric discretisation scheme applied to
the Abelian Chern-Simons theory}
\author{Samik Sen, Siddhartha Sen,  James C.\ Sexton and David H.\ Adams \\
\small{\it School of Mathematics, Trinity College, Dublin 2, Ireland} }
\maketitle

\begin{abstract}
We give a detailed general description of a recent geometrical
discretisation scheme and illustrate, by explicit numerical
calculation, the scheme's ability to capture topological features.
The scheme is applied to the Abelian Chern-Simons theory and leads,
after a necessary field doubling, to an expression for the discrete
partition function in terms of untwisted Reidemeister torsion and of
various triangulation dependent factors.  The discrete partition
function is evaluated computationally for various triangulations of
$S^3$ and of lens spaces. The results confirm that the discretisation
scheme is triangulation independent and coincides with the continuum
partition function.
\end{abstract}



\section{INTRODUCTION}

A very useful way to regularize a quantum field theory is provided by
the lattice formulation introduced by Wilson \cite{Wilson}.  (See for
example, \cite{creutz} for a detailed treatment).  However, this
formulation has difficulty in capturing topological features of a
field theory, for example, the topological theta-term in QCD.  It is
therefore of interest to investigate alternative discretisation
schemes.  In this paper we describe an alternative scheme which is
applicable to antisymmetric tensor field theories including Abelian
gauge theories and fermion field theory in the K\"ahler--Dirac
framework \cite{BJ}, and which is well-suited for capturing the
topological features of such theories. It is based on developing
analogies between the different types of fields and the way they
appear in a quantum field theory with a corresponding list of discrete
variables and operators.  The method is valid for any arbitrary
compact 3-manifold without boundary.  We illustrate the scheme by
applying it to the pure Abelian Chern-Simons gauge theory in 3
dimensions, and to a doubled version, the so-called Abelian BF gauge
theory. In the latter case the topological features of the theory are
completely reproduced by our discretisation scheme, even before taking
the continuum limit.  We illustrate this by explicit numerical
calculations of the discrete partition function when the spacetime is
$S^3$ or is a lens space, L(p,1).

The discretisation scheme involves several ingredients which are not
used in the standard Wilson lattice formulation.  These include a
triangulation of the spacetime ({\it i.e.} a decomposition into
hyper-tetrahedra rather then hyper-cubes) and a mathematical tool
called the Whitney map \cite{Whitney}. These have previously been used
to discretise field theories in \cite{albeverio,albevario2} where
various convergence results (e.g. convergence of the discrete action
to the continuum action) were established.  Triangulations of
spacetime are also used in discretisations of other quantum field
theories, for example in quantum gravity (see \cite{loll} for a
review).  The thrust of the present work is quite different.  We set
up the discretisation in such a way that the geometric structures of
the continuum field theory are mirrored by analogous structures in the
discrete formulation. As we will see, this requires a certain doubling
of fields. With this doubling the topological features of Abelian
Chern-Simons theory are completely captured by the formulation.  An
alternate scheme of discretisation proposed previously in
\cite{albevario2} which does not involve the doubling of fields fails
to capture the topological features of the Abelian Chern-Simons theory
as we will show numerically in Section~\ref{sec5}.  A mathematical
treatment of this discretisation scheme has been given in 
\cite{adth,adhep}. A short version of some of the results has been published in
\cite{adprl,sexton}.

Our aims in this paper are, firstly, to make the techniques and
results of \cite{adth,adhep} accessible to a wider audience, and
secondly, to demonstrate a practical numerical implementation of the
discretisation scheme.  Numerical implementations are developed for
Abelian Chern-Simons theory defined on the three dimensional sphere,
$S^3$, and on the lens spaces $L(p,1)$ for $p=1,2$ and $3$.  Our
discretisation scheme is the only one which numerically reproduces the
exact topological results for Abelian Chern-Simons theory \cite{only}.

This paper is organised as follows. In Section~\ref{sec2} the
discretisation scheme is described together with a summary of the
topological results used to set it up.  In Section~\ref{sec3} some
features of the Abelian Chern-Simons theory on general 3-manifolds are
reviewed. In Section~\ref{sec4} the discretisation scheme is applied
to this theory. An expression for the partition function of the
resulting discrete theory is derived in terms of the data specifying
the triangulation of the spacetime.  In Section~\ref{sec5} the
numerical evaluations of the partition functions corresponding to
these triangulations are presented. In Section~\ref{sec6} we summarise
our conclusions.


\section{THE DISCRETISATION SCHEME}
\label{sec2}

Let us start by considering a general quantum field theory defined on
an arbitrary manifold M of dimension D. Suppose the theory has fields
$\phi^p(\vec{x})$ where $\vec{x}\in M$, and where $\phi^p$ is a
$p$-form (antisymmetric tensor field of degree $p$ defined on $M$).
The Lagrangian for the system involves the fields, the Laplacian
operator, and possibly (as for the Chern-Simon theory) an
antisymmetrized first-order differential operator.

In differential geometric terms, the theory is constructed using the
following objects which are defined on the manifold $M$: $p$-forms
$\phi^p$ which are generalised antisymmetric tensor fields, the
exterior derivative $d:\phi^p \to \phi^{p+1}$, the Hodge star operator
$*:\phi^p \to \phi^{D-p}$, which is required to define scalar
products, and the wedge operator $\phi^p \wedge \phi^q = \phi^{p+q}$.

We want
to construct discrete analogues of these objects. We begin by summarizing
the basic properties of our operators of interest \cite{Eguchi}.
On a manifold M of dimension D, the operations ($\wedge, *, d$) on
p-forms, $\phi^p (p=0,\dots,D)$, satisfy the following: 
\begin{enumerate}
\item 
$\phi^p\wedge\phi^q = (-1)^{pq}\phi^q\wedge\phi^p$. 
\item 
$d(\phi^p\wedge\phi^q) = d\phi^p \wedge \phi^q + (-1)^p\phi^p\wedge d\phi^q$.
\item 
$*\phi^p = \phi^{D-p}$.
\item 
$* * = (-1)^{Dp+1}$.
\item 
$d^2=0$, $(d^*)^2=0$. 
\item 
$d^*=(-1)^{D(p+1)+1}*d*$, ($d^*$ is the adjoint of $d$).
\end{enumerate}
The following definitions will also be required
\begin{itemize}
\item 
The Laplacian on p-forms $\Delta_p = d_{p-1} d_p^{*}+d_{p+1}^{*}d_p$. 
\item 
The inner product $\langle \phi_p,{\phi_p}' \rangle = \int_M\phi_p\wedge *{\phi_p}'$.
\end{itemize}

A few examples might now be helpful.  Firstly, consider QED in 4
dimensions.  The gauge field $A=\phi^1$ is a 1-form. The
electromagnetic field is a 2-form given by $F=dA$.  The action for the
gauge field in QED is given by
\begin{equation}
S(A)=(F,F)= \langle dA, dA \rangle =\int_M dA\wedge *dA.
\end{equation}
Thus S involves the operators *, d and the wedge product.
Similarly for Abelian Chern-Simons theory  
the gauge field $A$ and electromagnetic field $F=dA$ are 1-forms and
2-forms respectively, as for QED.  The action for the theory is given
by
\begin{equation}
S(A)=\int_M A\wedge dA = \langle A, *dA \rangle
\end{equation}
where the spacetime M is 3-dimensional.
Note that in both cases we can think of the action $S(A)$ as a quadratic
functional of the gauge field $A$.

We would like to discretise the fields of $\phi^p$, the inner product
$\langle \cdot, \cdot \rangle$, and the operators $(\wedge, *, d)$
such that discrete analogues of their continuum interrelationships
hold.  To do this it is necessary to first introduce a few basic ideas
of discretisation.
We start by discretising the manifold $M$. This involves replacing
$M$ by a collection of discrete objects, known as simplices, glued
together. We need a few definitions \cite{NS}.

Firstly, for $p\geq 0$, a $p$-simplex $\sigma^{(p)} = [v_0,\dots,v_p]$ 
is defined to be the convex hull in some 
Euclidean space ${\mathbb{R}}^D$ of a set of $p+1$ points $v_0,v_1,\dots,v_p\in
{\mathbb{R}}^m$.  Here the vertices $v_i$ are required to span
a $p$-dimensional space.  This requirement will hold so long as the
equations, $\sum_{i=0}^p\lambda_i v_i = 0$ and $\sum_{i=0}^p\lambda_i=0$
admit only the trivial solution $\lambda_i=0$ for $i=0,\dots, p$ for 
$\lambda_i$ real.

A few examples might clarify the geometry. Consider
$\sigma^{(0)}=[v_0]$. This is a point or 0-simplex.  Next
$\sigma^{(1)}=[v_0, v_1]$ is a line segment or 1-simplex.  An
orientation can be assigned by the ordering of the vertices, in which
case $-\sigma^{(1)}= [v_1,v_0]$ for example.  The faces of a 1-simplex
are its vertices $[v_0]$ and $[v_1]$ which are 0-simplices.
$\sigma^{(2)}=[v_0, v_1, v_2]$ is a triangle or 2-simplex.  We note
that an even permutation of the vertices has the same orientation as
$\sigma^{(2)}$ while an odd permutation reverses it and will be
written as $-\sigma^{(2)}$.  The faces of a 2-simplex are its edges
$[v_0,v_1]$, $[v_1,v_2]$, and $[v_2,v_0]$.  Finally,
$\sigma^{(3)}=[v_0,v_1,v_2,v_3]$ is a tetrahedron or 3-simplex.  Its
faces are the four triangles $[v_0,v_1,v_2], [v_0,v_2,v_3],
[v_0,v_1,v_3]$ and $ [v_1,v_2,v_3]$ which bound it.

\begin{figure}
\begin{center}
\includegraphics[width=200pt]{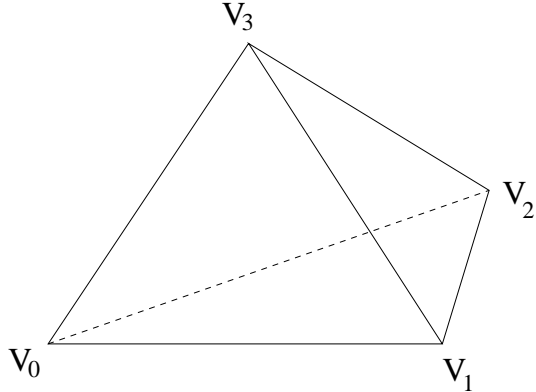}
\caption{A 3-simplex.}
\label{fig1}
\end{center}
\end{figure}

An important feature of our discretisation scheme is that the original
simplices are subdivided by using simplex barycentres.  Geometrically
the {\em barycentre} of a $p$-simplex, $\sigma^{(p)}$, is the point
which represents its ``center of mass''. We denote the barycentre of
$\sigma^{(p)} = [v_0, \dots, v_p]$ as the point
\begin{equation}
\hat{\sigma}^{(p)} = \frac{1}{p+1}\sum_{i=0}^p v_i.
\end{equation}
As an example, the barycentre of $\sigma^{(1)}=[v_0,v_1]$ is the
midpoint of the line segment which joins the vertices $v_0$ and $v_1$.

We can now describe a particular way that a given manifold $M$ can be 
discretised. Let $S$ be a collection of simplices $\{\sigma^{(n)}_i\}$, 
$n=0,1,\dots,D$, with the property that the faces of the simplices which 
belong to $S$ also belong to it. The elements of $S$ glued together in 
the following way is known as a simplicial complex\cite{NS,HY}: 
\begin{enumerate}
\item $\sigma_i^{(n)} \cap \sigma_j^{(k)} = 0$ if
$\sigma_i^{(n)}$, $\sigma_j^{(k)}$ have no common face. \\
\item $\sigma_i^{(n)} \cap \sigma_j^{(k)} \neq 0$ if
$\sigma_i^{(n)}$, $\sigma_j^{(k)}$ have precisely one face in common,
along which they are glued together. \\
\end{enumerate}
In many cases of interest(including all 3-manifolds and all differentiable
manifolds \cite{HY}), $M$ can be replaced by a complex $K$ which it is
topologically equivalent to. $K$ is then said to be a triangulation of $M$(Note this 
triangulation is not unique).  \noindent
In this way of discretising M, the building blocks are zero, one, $\dots$, 
D-dimensional objects, all of which are simplices e.g.generalised 
oriented tetrahedra.

We now observe that the same manifold can be discretised in many
different ways.  In the discretisation described, we used simplices.
We could just as well have used generalised oriented cubes.

\begin{figure}[h]
\begin{center}
\includegraphics[width=170pt]{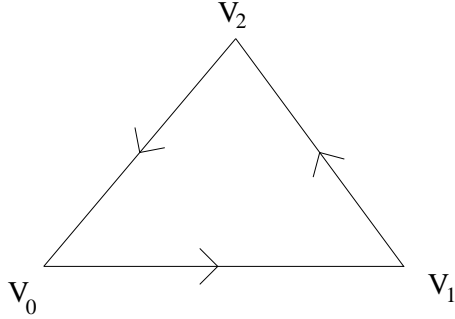}
\caption{Triangulation $K$ of a disc.}
\label{fig2}
\end{center}
\end{figure}

There is another method of discretising a manifold which is the dual
of the simplicial discretisation just described.  It associates with a
simplicial complex $K$ a dual complex $\hat{K}$. We
proceed to describe this construction.  We will see that the basic
objects of the dual complex $\hat{K}$ are again zero, one, two,
$\dots$, D dimensional objects, but this time they are no longer
simplices. We illustrate the method by considering a manifold, $M$,
which is a disc. This is a manifold with a boundary. We triangulate
this by the simplicial complex, $K$, shown in Figure~\ref{fig2}.

Now consider the barycentres of the building blocks of the simplicial
complex $K$. We have the following list shown in Table~\ref{tab1}.  
\begin{table}
\begin{center}
\begin{tabular}{ccc}
Geometrical object in K & dimension & Corresponding barycentre \\
\hline
$\sigma^{(0)}_1=[v_0]$         & 0 & $\hat{\sigma}^{(0)}_1 = v_0$, \\
$\sigma^{(0)}_2=[v_1]$         & 0 & $\hat{\sigma}^{(0)}_2 = v_0$, \\
$\sigma^{(0)}_3=[v_2]$         & 0 & $\hat{\sigma}^{(0)}_3 = v_0$, \\
$\sigma^{(1)}_1=[v_0,v_1]$     & 1 & $\hat{\sigma}^{(1)}_1=\frac{1}{2}(v_0+v_1)=v_3$ \\
$\sigma^{(1)}_2=[v_1,v_2]$     & 1 & $\hat{\sigma}^{(1)}_1=\frac{1}{2}(v_1+v_2)=v_4$ \\
$\sigma^{(1)}_3=[v_2,v_0]$     & 1 & $\hat{\sigma}^{(1)}_1=\frac{1}{2}(v_2+v_0)=v_5$ \\
$\sigma^{(2)}_1=[v_0,v_1,v_2]$ & 2 & $\hat{\sigma}^{(2)}_1=\frac{1}{3}(v_0+v_1+v_2)=v_6$ \\
\end{tabular}
\end{center}
\caption{Elements of the simplicial complex $K$ and their barycentres.}
\label{tab1}
\end{table}

\begin{figure}[h]
\begin{center}
\includegraphics[width=200pt]{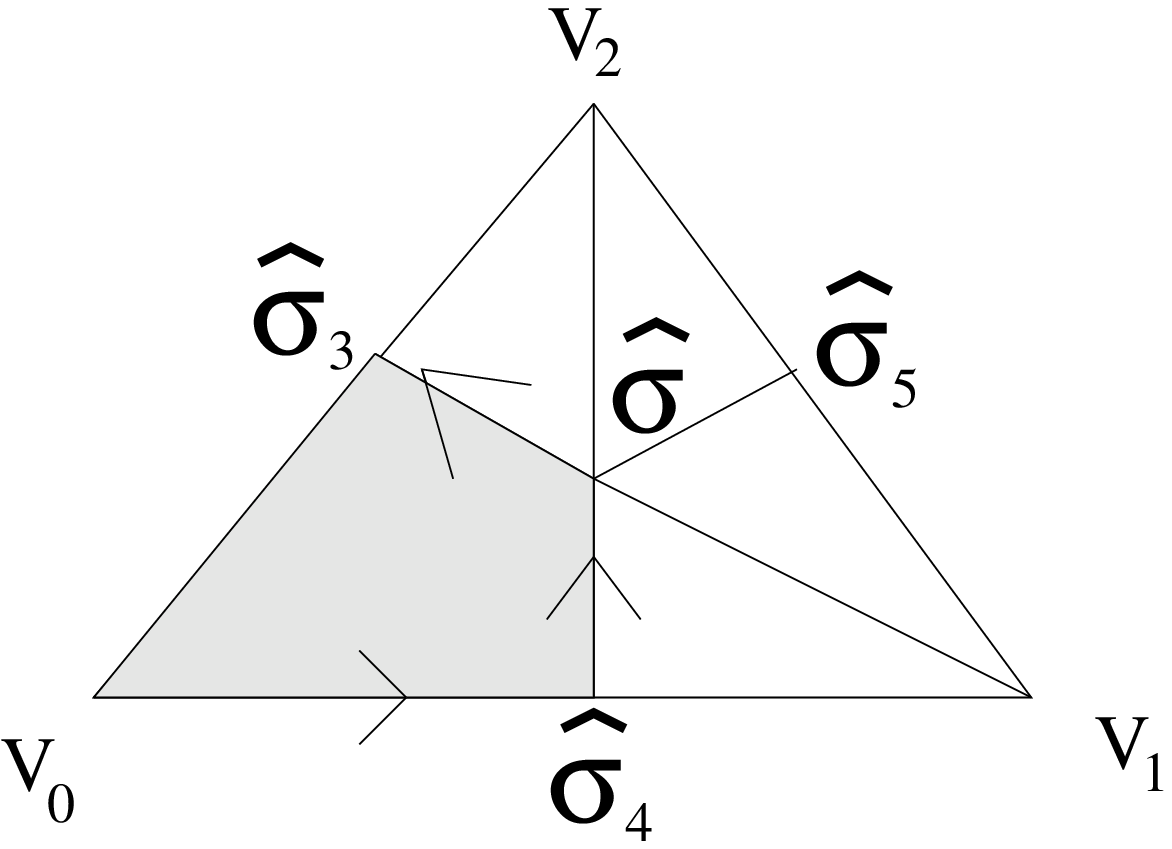}
\caption{Barycentres of $K$.}
\label{fig3}
\end{center}
\end{figure}

Pictorially the simplicial complex $K$ with its barycentres is shown in 
Figure~\ref{fig3}.

We can now construct the dual description of a triangulation K.
Geometrically this utilises a discrete analogue of the Hodge
${\mathbb{*}}$ operator. Recall the * operator maps a p-form to a $(D-p)$-form
, where D is the dimension of the manifold M on which the p-form is defined.
In the dual geometrical decomposition of the manifold, we want to set
up a correspondence between a p-dimensional object and a $(D-p)$ dimensional 
object. 

This is done as follows. We first construct $(D-p)$ dimensional
objects whose vertices are barycentres of a sequence of successively higher
dimensional simplices, where each simplex is a face of the following one.
In other words $(D-p)$ dimensional objects of the form 
$\{\hat{\sigma}_p, \hat{\sigma}_{p+1}, \dots, \hat{\sigma}_D\}$, 
where $\sigma_n$ is a face of $\sigma_{n+1}$. The orientation of
these are set so as to be compatible with the manifold.
Joining these objects together gives us the dual of $\sigma_p$.

\begin{figure}[h]
\begin{center}
\includegraphics[width=200pt]{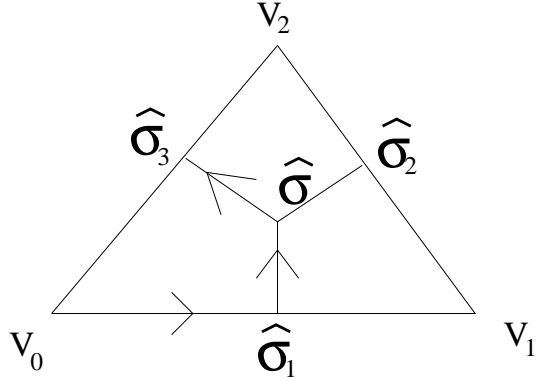}
\caption{Dual complex $\hat{K}$.}
\label{fig4}
\end{center}
\end{figure}

Thus for instance the map * acts on $[v_0]$ as follows.
\[ *_K:[v_0]\to\epsilon_{01}[\hat{v_0},\hat{\sigma_1},
\hat{\sigma}]\cup[\hat{v_0},\hat{\sigma_3},\hat{\sigma}]\epsilon_{03}.\]
Where the orientation of each of the small triangles has to be coherent with 
the orientation of the original triangle. This is shown in fig. 3
and leads to mapping $[v_0]$ to the shaded two dimensional region.
The orientations of the simplices are specified by arrows in the figure.
Coherence of orientation means, for example, that the arrow of an edge
agrees with the arrow of the triangle to which it belongs.
Next we consider $[v_0,v_1]$. This is a 1-simplex and is to be mapped
to a (2-1)=1 dimensional object. The map is defined as
\[ *_K:[v_0,v_1]\to [\hat{\sigma_1},\hat{\sigma}].\]
Again the orientation of $[\hat{\sigma_1},\hat{\sigma}]$ has to be
coherent with the orientation of the triangles already introduced
when the map for $[v_0]$ was considered. Similarly
\begin{eqnarray*}
*_K:[v_1,v_2] & \to & [\hat{\sigma_2},\hat{\sigma}], \\
*_K:[v_2,v_0] & \to & [\hat{\sigma_3},\hat{\sigma}], \\
\end{eqnarray*}
and finally
\begin{eqnarray*}
*_K:[v_0,v_1,v_2] &\to [\hat{\sigma}].
\end{eqnarray*}
We then have the alternate discretisation $\hat{K}$ for M shown 
in fig. \ref{fig4}.


Note when two edges are glued together they must have opposite orientations.
We can now give the general rule for mapping an n-simplex
$\sigma_n = [v_0,\dots,v_n]$ to a ($D-n)$ dimensional object (
(D-n) cell) as follows:
We think of $\sigma_n$ as an element of a simplicial complex K. We have
\[*_K:[v_0,\dots,v_n]\to\cup[\hat{\sigma_n},
\hat{\sigma_{n+1}},\dots,\hat{\sigma_D}], \]
where $\hat{\sigma}_{n+1}$ is the barycentre of an (n+1)-simplex which 
has $\sigma_n$ as a face.
$\hat{\sigma}_{n+2}$ is the barycentre of an (n+2)-simplex 
which has $\sigma_{n+1}$ as a face and so on. These objects have to
be coherently oriented with respect to $[v_0,\dots,v_n]$. The set
of these cells constitutes the dual space $\hat{K}$ of K.

By this procedure we claim, a discrete version of the Hodge star operation 
* has been constructed. Let us explain. The Hodge * operator involves forms.
It maps p-forms in D dimensions to a (D-p)-form. The $*_K$ map 
involves not forms but geometrical objects. There is a 
simple correspondence relation between these two cases. 
Given a p-form, $\phi_p$, and a p-dimensional geometrical space, $\Sigma_p$, 
the p-form can be integrated over $\Sigma_p$ to give a number. 
Thus $\Sigma_p$ and $\phi_p$ are objects that can be paired.
We can write this as a pairing
\[ (\phi_p, \Sigma_p) =\int_{\Sigma_p}\phi_p .\]
In order to proceed, we need to introduce some more structure.
We start by associating with a simplicial complex K, containing
$\{\sigma_p^i\}$ $(i=1,\dots,K_p;p=0,\dots,D)$ a vector space
consisting of finite linear combinations over the reals of the
p-simplices it contains. This vector space is known as the space of
p-chains,$C_p(K)$.
For two elements $\sigma_p^i, \sigma_p^j \in C_p(K)$, a scalar
product $(\sigma_p^i,\sigma_p^j)=\delta_j^i$ can be introduced.
An oriented p-simplex changes sign under a change of orientation i.e.
if $\sigma_p=[v_0,\dots,v_p]$ and $\tau$ is a permutation of the 
indices $[0,\dots,p]$, then $[v_{\tau(0)},\dots,v_{\tau(p)}]=
(-1)^\tau[v_0,\dots,v_p]$, with $\tau$ denoting the number of transpositions
needed to bring $[v_{\tau(0)},\dots,v_{\tau(p)}]$ to the order
 $[v_0,\dots,v_p]$.

Given the vector space $C_p(K)$, the boundary operator $\partial^K$ can
be defined as
\[ \partial^K:C_p(K)\to C_{p-1}(K). \]
It is the linear operator which maps an
oriented p simplex $\sigma^{(p)}$ to the sum of its (p-1) faces
with orientation induced by the orientation of $\sigma^p$.
If $\sigma^p=[v_0,\dots,v_p]$, then 
\[ \partial\sigma^p=\sum_{i=0}^p (-1)^i[v_o,\dots,\hat{v_i},\dots v_p],\]
where $[v_o,\dots,\hat{v_i},\dots, v_p]$ means that the vertex $v_i$
has been omitted from $\sigma^p$ to produce the face
``opposite'' to it.

Given that $C_p(K)$ is a vector space, it is possible to define
a dual vector space $C^p(K)$, consisting of 
dual objects known as cochains;  that is we can take an element
of $C_p(K)$ and an element $C^p(K)$ to form a real number. 
Since the space $C_p(K)$ has
a scalar product, namely if $\sigma_p^i, \sigma_p^j \in C_p(K)$
then $(\sigma_p^i,\sigma_p^j)=\delta_{ij}$. 
We can use the scalar product to identify $C_p(K)\equiv C^p(K)$,
so 
that we can consider oriented p-simplices as elements of $C^p(K)$ as well
as $C_p(K)$. We can write our boundary operation as
\[ ([v_0,\dots,\hat{v_i},\dots,v_p],\partial^K[v_0,\dots,v_p])=(-1)^i.\]
This suggests introducing the adjoint operation $d_K$ defined as
\[
(d^K[v_0,\dots,\hat{v_i},\dots,v_p],[v_0,\dots,v_p])=
([v_0,\dots,\hat{v_i},\dots,v_p],\partial^K[v_0,\dots,v_p]).
 \]
This is the coboundary operator which maps $C_p(K)\to C_{p+1}(K)$. 

Indeed we have
\[ d^K[v_0,\dots,v_p]=\sum_v[v,v_0,\dots,v_p], \]
where the sum is over all vertices $v$ such that
$[v,v_0,\dots,v_p]$ is a (p+1) simplex.

The boundary operators $\partial_K$ and the coboundary
operator $d_K$ have the property $\partial_K\partial_K=d_Kd_K=0$.
Furthermore,
\begin{eqnarray*}
d_K:C_p &\to & C_{p+1}, \\
 \partial_K:C_p &\to & C_{p-1}.
\end{eqnarray*}
These operators are the discrete analogues of the operators
$d$ and $(-1)^{D(p+1)+1} *d*=d^*$ which act on forms.

These operators could be defined only when a scalar
product(``metric'') was introduced in the vector space $C_p$'s.
At this stage we have a discrete geometrical analogue of 
$d$, $d^*$ and $*$. We have also commented on the fact that the operation
$*$ maps simplices into dual cells i.e. not simplices.  If the original 
simplicial system is described in terms of the union of the vector spaces of all
p-chains then the space into which elements of the vector space
are mapped by * is not contained within this space, unlike the
situation for the Hodge star operation on forms. We will see that this
difference leads inevitably to a doubling of the fields when
discretisation, preserving topological structures, is attempted.

We now need a way to relate a p-chain to a p-form. This 
together with a construction which linearly maps p-forms to p-simplices
will allow us to translate expressions in continuum QFT to a corresponding
discrete geometrical objects. We start with the construction of the linear
maps from p-chains to p-forms due to Whitney \cite{Whitney}.

 In order to define this map, we need to introduce
barycentric coordinates associated with a given p-simplex $\sigma^p$. 
Regarding $\sigma^p$ as an element of some ${\mathbb{R}}^N$, we introduce
a set of real numbers $(\mu_0,\dots,\mu_p)$ with the property
\begin{eqnarray*}
\mu_i & \geq & 0, \\
\sum_i \mu_i &=& 1.
\end{eqnarray*}
A point $x\in \sigma^p$ can be written in terms of the vertices of 
$\sigma^p$ and these real numbers as
\[ x=\sum_{i=0}^p \mu_i v_i. \]
Note if any set of $\mu_i=0$ then the vector $x$ lies on a face
of $\sigma^p$. One can think of $x$ as the position of the
center of mass of a collection of masses $(\mu_0,\dots,\mu_p)$ located
on the vertices $(v_0.\dots,v_p)$ respectively.
Setting $\mu_i=0$ for instance means the center of mass will be 
in the face opposite the vertex $v_i$. The Whitney map can now
be defined. We have
\[ W^K:C^p(K)\to\Phi^p(K), \]
where $\Phi^p(K)$ is a p-form. If $\sigma^p\in C^p(K)$ then
\[ W[\sigma^p]=p!\sum^p_{i=0}(-1)^i\mu_i d\mu_0\wedge\dots
\hat{d\mu_i}\wedge\dots d\mu_p, \]
where $\hat{d\mu_i}$ means this term is missing, and $(\mu_0,\dots,
\mu_p)$ are the barycentric coordinate functions of $\sigma^p$.

We next construct the linear map from p-forms to p-chains. This is 
known as the de Rham map. We have
\[ A^K:\Phi^p(K)\to C^p(K), \]
defined by
\[ <A^K(\Phi^p),\sigma^p> = \int_{\sigma^p}\Phi^p, \]
for each oriented p-simplex $\in K$. 

%

A discrete version of the wedge product can also be 
defined using the Whitney and de Rham maps such that 
$\wedge^K:C^p(K)\times C^q(K)\to C^{p+q}(K)$ as follows:
\[ x \wedge^K y = A^K(W^K(x)\wedge W^K(y)).\]
It has many of the properties of the continuous wedge product in that 
it is skewsymmetric and obeys the Leibniz rule but it is nonassociative.

At this stage we have introduced all the building blocks necessary
to discretise a system preserving geometrical structures. We summarize
the properties of the maps introduced in the form of a 
theorem \cite{Whitney}: 
\begin{enumerate}
\item $A^K W^K$ = Identity. \\
\item $dW^K = W^K d^K$, where $ d:\phi^p \to \phi^{p+1}$. \\
\item $\int_{\mid\beta\mid}W^K(\alpha)=<\alpha,\beta>$, $\alpha, \beta
\in K$\\
\item $ d^K A^K = A^K d $. \\
\end{enumerate}

\noindent
This theorem shows how $d^K$ can be considered as the discrete analogue of
$d$. We now show how $*^K$ can be considered as discrete analogue of
$*$. For this we need barycentric subdivision.

 We recall that given a simplicial complex
$\{\sigma^p_i\}$, i=1,$\dots,K_p;p=0,\dots,D$. A set of points(vertices)
could be assigned to each simplex, namely $\hat{\sigma_i^p}$. These
are the barycentres. These vertices, regarded as vertices of a simplex,
subdivide the original simplices to give a finer triangulation of the
original manifold. This is a barycentric subdivision map $BK$. 
Clearly the procedure
can be repeated to give finer and finer subdivisions in which the 
simplices become ``smaller''. The procedure is illustrated for a triangle
in Fig. 5.

\begin{figure}
\begin{center}
\includegraphics[width=350pt,height=100pt]{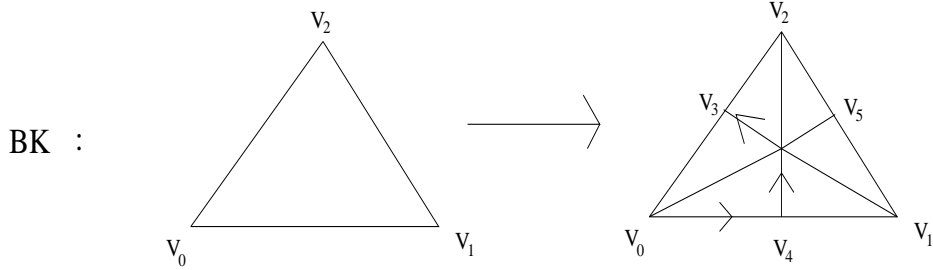}
\caption{Barycentric subdivision of $K$.}
\label{fig5}
\end{center}
\end{figure}


Note that all the barycentres are present as vertices
of the barycentric subdivision and that $*^K$ acting on simplices
belonging to the simplicial complex, $K$,  associated with $[v_0,v_1,v_2]$
leads to objects which are not, in general, simplices of $[v_0,v_1,v_2]$
but belong to a different space $\hat{K}$.
However both $K$ and $\hat{K}$ are contained in the barycentric subdivision 
$B[v_0,v_1,v_2]$.
This is a crucial observation. In order to construct 
the star map, two geometrically distinct spaces were introduced. The 
original simplicial decomposition $K$ with its associated set of p-chains
$C^p(K)$ and the dual cell decomposition $\hat{K}$ with its associated 
set of p-chains $C^p(\hat{K})$. 
These spaces are distinct. However both belong to the
first barycentric subdivision of $K$. This allows the use the $*^K$
operation if we think of $K$ and $\hat{K}$ as elements of $BK$.

We proceed as follows. Let BK and $\hat{K}$
denote the barycentric subdivision and dual triangulation, and
\[ *^K:C^p(K)\to C^{n-p}(\hat{K}). \]
However $C^p(K)$ and $C^p(\hat{K})$ are both contained in 
$C^p(BK)$ as we have seen. Let 
\[ W^{BK}:C^p(BK)\to\phi^p(M),\]
denote the Whitney map. Then we have for $x\in C^p(K), y 
\in C^{n-p-1}(\hat{K})$\cite{adhep}:
\begin{enumerate}
\item $<*^K x,y > = \frac{(n+1)!}{p!(n-p)!}\int_M W^{BK}(Bx)\wedge W^{BK}(By).
\\ \\ 
<*^{\hat{K}} y,x > = \frac{(n+1)!}{p!(n-p)!}\int_M 
W^{BK}(By)\wedge W^{BK}(Bx). $\\ \\
\item $\partial^K = (-1)^{np+1}*^{\hat{K}}d^{\hat{K}}*^K$ on $C^p(K)$.\\
$\partial^{\hat{K}} = (-1)^{nq+1}*^Kd^K*^{\hat{K}}$ on $C^q(\hat{K})$.\\
\end{enumerate}

\noindent
These are the discrete analogues of the interrelationships between $d, d^*,
*$ and $<\cdot,\cdot>$ in the continuum.

Note $K\neq \hat{K}$ and that properties of $\partial_K, d_K$
analogous to those for differential forms only hold if 
$K$ , $\hat{K}$ are both regarded as elements of BK. 
This feature of the discretisation method is, as we shall see, crucial 
if we want to preserve topological properties of the original system. 
If a discretisation method is introduced without the $*$ operation in it then
as we shall see in section~\ref{sec5} the topological 
properties of the partition function for the Abelian Chern-Simons gauge theory 
do not hold. 

We proceed to apply these ideas to the Abelian Chern-Simons
gauge theory on a compact three manifold $M$. First we summarize properties
of the continuum field theory.

\section{Schwarz's topological field theory and the Ray-Singer torsion}
\label{sec3}
We begin our treatment of continuum field theory by describing Schwarz's method 
for evaluating the partition function of
the Chern-Simons gauge theory on a three-manifold $M$ \cite{Schwarz}. 
We assume that
the first real homology group ( to be defined shortly)  of the manifold 
vanishes; this is done for the sake of simplicity. Schwarz's method is 
applicable for arbitrary compact 3-manifolds
without boundary.
Such manifolds are called ``homology 3-spheres''.  The main example we have in mind
are the 3-sphere $S^3$ and the lens spaces $L(p,1)$, p=1, 2,
$\dots$( for a definition and the basic properties of lens spaces see 
\cite{Munkres,lens}).
The fields of the theory are the 1-forms on M i.e. $\omega \in \Omega^1(M)$.
(In terms of a local coordinate system $(X^\mu)$ on M we have $\omega(x)=
\omega_\mu(x)dx^\mu$.) The action is 
\begin{eqnarray}
S(\omega) = \int_M \omega\wedge d_1\omega = \int_Mdx^1dx^2dx^3
\epsilon^{\mu\nu\rho}\omega_\mu\partial_\nu\omega_\rho.
\label{eq:29}
\end{eqnarray}
Here and in the following $\Omega^{q}(M)$ denotes the space
of $q$-forms on M (i.e. the antisymmetric tensor fields of degree
q) and $d_q:\Omega^q(M)\to \Omega^{q+1}(M)$ i.e. the exterior
derivative. It has the property $d_qd_{q-1}=0$ so ${\rm Im}(d_{q-1})
\subset {\rm Ker}(d_q)$, where $Im(d_{q-1})$ is the image of the operator
$d_{q-1}$ while $Ker(d_q)$ is the null space of the operator $d_q$ 
and the cohomology spaces $H^q(M)$ are
defined by
\[ H^q(M)={\rm Ker}(d_q)/{\rm Im}(d_{q-1}). \]
The $H^q(M)$ are Abelian groups which contain topological information about the
manifold. The vanishing of $H^1(M)$, for instance, holds if the manifold is
simply connected that is any loop in $M$ can be smoothly deformed to any
other loop in $M$ \cite{NS}.
Note that $\Omega^0(M)$ is the space of functions on M and
since $d_0$ is the derivative ${\rm Ker}(d_0)$ consists of the constant
functions i.e.
\[ H^0(M)={\rm Ker}(d_0)/{0} = {\rm Ker}(d_0) = {\mathbb{R}}. \]
Our requirement on M that $H^1(M)=0$ implies that 
\[ {\rm Im}(d_0) = {\rm Ker}(d_1). \]
A choice of metric on M determines an inner product in the spaces
$\Omega^q(M)$ and allows the action (\ref{eq:29}) to be written as
\begin{eqnarray}
S(\omega) = \lambda <\omega, (*d_1)\omega>, 
\label{eq:30}
\end{eqnarray}
where * is the Hodge star operator. ( See \cite{Eguchi} for background
on this and other differential-geometric constructions.)
In order to evaluate the partition function of this action
by Schwarz's method requires the introduction of the resolvent for $S(\omega)$.
The partition function is defined as
\begin{eqnarray*}
 Z(\lambda)  &=& N\int d\omega e^{iS(\omega)}. \\
\end{eqnarray*}
The main problem in evaluating $Z(\lambda)$ is to properly deal with the zeroes of 
$S(\omega)$. These zero modes contain topological information regarding the 
manifold as the space of zero modes is given by ${\rm Ker}(d_1)$ and hence 
should not be discarded.
Schwarz introduced an algebraic method (the resolvent method) for dealing with this problem.
Although it is only 
valid for $S(\omega)$'s which are quadratic in $\omega$, it can be used to analyse 
$S(\omega)$'s constructed on arbitrary compact manifolds without boundary. For systems of 
this type Schwarz's method is an algebraic analogue of the problem of gauge fixing.
The advantage of the resolvent method is that it can be easily extended to deal with
the process of discretisation as we will show.

The resolvent is defined to be the following chain of maps 
\begin{eqnarray}
0 \to {\mathbb{R}} \to^{\phi_0}\Omega^0(M) \to^{d_0}{\rm Im}(d_0)=
{\rm Ker}(d_1) \to {\rm Ker}(S) \to 0. 
\label{eq:31}
\end{eqnarray}

These chain of maps form an exact sequence, that is, the image of a map
is the kernal of the map which follows.
With the help of the resolvent, Schwarz was able to show that the partition
function for the theory was given by \cite{Schwarz} 
\begin{equation}
Z(\lambda)=e^{-\frac{i\pi}{4}\iota}(\frac{\lambda}{\pi})^{-\frac{\zeta}{2}}
{\det} '((*d_1)^2)^{-\frac{1}{4}}{\det} '(d^*_0 d_0)^{\frac{1}{2}}
\det(\phi_0^*\phi_0)^{-\frac{1}{2}}, 
\label{eq:32}
\end{equation}
where $\iota$ is a non-topological geometry dependent function and $\zeta$ as shown in
\cite{adhep} is given by
\[ \zeta={\rm dim }H^0(M) - {\rm dim }H^1(M). \]
$\iota$ is part of a phase factor and thus the absolute value of the partition function 
$Z(\lambda)$ is a topological quantity. This will be our main concern \cite{adprl}.

For completeness we give a quick proof of this result, ignoring phase
factors and constants.

Introducing a metric in the space of $\omega$'s allows us to write
\begin{eqnarray*}
Z(\lambda)  &=& N\int_{{\rm Ker } d_1 \oplus ({\rm Ker } d_1)^\perp} 
d\omega e^{iS(\omega)}, \\
&=&\mbox{Vol} ({\rm Ker } d_1)\cdot(\det d_1^*d_1)^{-\frac{1}{4}}N.
\end{eqnarray*}
We proceed to rewrite Vol(${\rm Ker } d_1$) using the exact sequence associated
with ${\rm Ker } d_1$ and the manifold $M$. This procedure gives an expression
for the partition function $Z$ containing information about the 
spaces ${\rm Ker } d_1$ and $({\rm Ker } d_1)^\perp$. Simply dropping 
Vol(Ker$d_1$) leads to a loss of information.
We have Vol(${\rm Ker } S$)=Vol(${\rm Ker } d_1$) = Vol(${\rm Im }  d_0$) 
by assumption (if $H_1(M)$ is non-trivial, this equation has to be modified 
\cite{adth}). Also,

\begin{eqnarray*}
d_0\mid_{({\rm Ker } d_0)^\perp} : ({\rm Ker } d_0)^\perp &\to& ({\rm Im }  d_0) \\
\Rightarrow \mbox{Vol}({\rm Im }  d_0) &=& \mid {\det} ' d_0 \mid \mbox{Vol} 
({\rm Ker } d_0)^\perp.
\end{eqnarray*}
\noindent
Note
\begin{eqnarray*}
\Omega^0 &=& {\rm Ker } d_0 \oplus ({\rm Ker } d_0)^\perp \\
\end{eqnarray*}
\noindent
and
\begin{eqnarray*}
\phi_0:{\cal H}_0 &\to& H_0 \\
\mbox{Vol}(H_0) &=&
\mid \det \phi_0 \mid
\mbox{Vol}({\cal H}_0).
\end{eqnarray*}

\noindent
where ${{\cal H}}_{\rm 0}$ represents the space of harmonic $0$-forms. Note harmonic
$p$-forms are solutions of $(d^*d + d d^*) \phi_p=0$. In this space the Hodge star
operator is present and hence a scalar product and volume can be defined. The map
$\phi_0$ introduced relates the space of harmonic $0$-forms to the space of de Rham
cohomology $H_0$. By a theorem of Hodge this space of harmonic $p$-forms is
isomorphic to the space of $H_p$\cite{Eguchi}. The space of de Rham cohomology does not have a 
metric 
and hence we define the volume in this space with the help of the map $\phi_0$.
Therefore,
\begin{eqnarray*}
\mbox{Vol}({\rm Ker } d_0)^\perp & =& \mbox{Vol}(\Omega_0)[\mbox{Vol}({\rm Ker } d_0)]^{-1},
\\
&=&\mbox{Vol}(\Omega_0)(\mbox{Vol}( H_0))^{-1}, \\
&=&(\mbox{Vol}(\Omega_0))(\det \phi_0)^{-1}(\mbox{Vol}({\cal H}_0))^{-1}.
\end{eqnarray*}
So that finally we get
\[ \mbox{Vol}({\rm Ker } S) = \mid {\det} ' d_o \mid \mid \det \phi_0 \mid^{-1}
\mbox{Vol}(\Omega_0)(\mbox{Vol}{\cal H_0})^{-1}. \]
Choosing 
\[ N\mbox{Vol}(\Omega_0)(\mbox{Vol} {\cal H_0})^{-1} = 1, \]
we get
\[ Z=\mbox{Vol}({\rm Ker } S)(\det d_1^*d_1)^{-\frac{1}{4}} = 
\mid {\det} ' d_0\mid \mid \det d_1d^*\mid^{-\frac{1}{4}}
\mid \det \phi_0\mid^{-1}. \]
A more careful calculation gives the determinant in (\ref{eq:32}). 
The quantities $\iota$ and $\zeta$ in (\ref{eq:32}) also 
need to be zeta-regularised - this was done in \cite{adhep}, where it was shown that the regularised $\zeta$ is given by 
\[ \zeta={\rm dim }H^0(M) - {\rm dim }H^1(M). \]
So in the present case, where $H^0(M)\equiv{\mathbb{R}}$ and $H^1(M)=0$, 
we have 
\begin{eqnarray}
\zeta=1-0=1.
\label{eq:33}
\end{eqnarray}
Using the formulae $d_1^*=*d_1*$ and $**=1$ (modulo a possible sign), we 
get $d_1=*d_1^**$ and therefore $(*d_1)^2=*d_1*d_1=d_1^*d_1$, which gives
\begin{eqnarray}
{\det} '((*d_1)^2) = {\det} '(d_1^*d_1). 
\label{eq:34}
\end{eqnarray}
Substituting (\ref{eq:33}) and (\ref{eq:34}) in (\ref{eq:31}) we get
\begin{eqnarray}
Z(\lambda)=e^{-\frac{i\pi}{4}\iota}(\frac{\lambda}{\pi})^{-\frac{\zeta}{2}}
{\det} '(d_1^*d_1)^{-\frac{1}{4}}{\det} '(d^*_0 d_0)^{\frac{1}{2}}
\det(\phi_0^*\phi_0)^{-\frac{1}{2}}. 
\label{eq:35}
\end{eqnarray}
We now rewrite the product of determinants in (\ref{eq:35})
in terms of the Ray-Singer torsion \cite{RS} of M. Since the Hodge
star operator * is unitary with $**=1$ and $d_0^*=*d_2*$
(modulo a possible sign) we have
\begin{eqnarray*}
{\det} '(d^*_0d_0) &=&{\det} '(*d_2*d_0) = {\det} '(*(*d_2*d_0)*),\\
&=&{\det} '(d_2*d_0*) ={\det} '(d_2d_2^*), \\
&=&{\det} '(d_2^*d_2).
\end{eqnarray*}
It follows that
\begin{eqnarray}
{\det} '(d_0^*d_0)^{\frac{1}{2}}{\det} '(d_1^*d_1)^{-\frac{1}{4}}
=({\det} '(d_0^*d_0)^{\frac{1}{2}}\det(d_1^*d_1)^{-\frac{1}{2}}
{\det} '(d_2^*d_2)^{\frac{1}{2}})^{\frac{1}{2}}.
\label{eq:36}
\end{eqnarray}
It is possible to rewrite $\det(\phi^*_0\phi_0)$ using a standard result of 
manifold theory in a different form(see \cite{Spivak}). We start by
noting that the integration map
\[ \int_M: H^3(M) \to {\mathbb{R}} \]
is an isomorphism, i.e. for each $r\in {\mathbb{R}} $ there is a unique
class $[\alpha]\in H^3(M)$ such that $\int\alpha=r$. (Note that
the integration map is well defined on $H^3(M)$ since  $\int_M \alpha
+ d\beta = \int_M \alpha$ i.e. $\int_M d\beta=0$ by Stokes
theorem.) Also from the definition
\[ H^3(M) = \Omega^3(M)/{\rm Im}(d_2) = ({\rm Im}(d_2)\oplus{\rm Im}(d_2)^
\perp)/{\rm Im}(d_2) = {\rm Im}(d_2)^\perp, \]
it follows that the map given by 
\begin{eqnarray}
{\rm Im}(d_2)^\perp \to {\mathbb{R}}
\label{eq:37}
\end{eqnarray}
is also an isomorphism. Now define the map
\[ \phi_3:{\mathbb{R}} \to {\rm Im}(d_2)^\perp \]
to be the inverse of (\ref{eq:37}). Then using the properties of the
Hodge star operator it can be shown that
\[ \det(\phi_3^*\phi_3)=\det(\phi_0^*\phi_0)^{-1}. \]
It follows that
\begin{eqnarray}
\det(\phi_0^*\phi_0)^{-\frac{1}{2}}=
(\det(\phi_0^*\phi_0)^{-\frac{1}{2}}\det(\phi_3^*\phi_3)^{\frac{1}{2}})
^{\frac{1}{2}}.
\label{eq:38}
\end{eqnarray}
Substituting (\ref{eq:38}) and (\ref{eq:36}) in (\ref{eq:35}) we get
\begin{equation}
\mid Z(\lambda) \mid = (\frac{\lambda}{\pi})^{-\frac{1}{2}}\tau_{RS}(M)^{
\frac{1}{2}},
\label{eq:39}
\end{equation}
where
\begin{equation}
\tau_{RS}(M)=\det(\phi_0^*\phi_0)^{-\frac{1}{2}}
\det(\phi_3^*\phi_3)^{\frac{1}{2}}{\det} '(d^*_0d_0)^{\frac{1}{2}}
{\det} '(d_1^*d_1)^{-\frac{1}{2}}{\det} '(d^*_2d_2) .
\label{eq:40}
\end{equation}
This quantity $\tau_{RS}(M)$ is the Ray-Singer
torsion of M \cite{RS}. It is a topological invariant of M i.e. is independent 
of the metric of M. Thus the modulus
$\mid Z(\lambda)\mid$ of the partition function, given by (\ref{eq:38})-(\ref{eq:39}), is a
topological invariant.

We are now ready to construct a discrete version of the preceding 
topological field theory which reproduces the continuum expression for the
partition function where subdivision invariance is the discrete property 
corresponding to topological invariance. 
We will see that in order to do this it is crucial 
that there is an analogue of
the Hodge star operator in the discrete theory. As we will see in the
next section, this requires a field doubling. Therefore we consider a 
doubled version of the preceding theory, with the fields $\omega_1$ and
$\omega_2$ in $\Omega^1(M)$ and with the action functional (\ref{eq:30}) changed by
\begin{equation}
S(\omega) = \lambda<w,(*d_1)w> \to \tilde{S}(\omega_1,\omega_2) = 
\lambda < ({}^{\omega_1}_{\omega_2}), ({}^{\,\, 0 \,\,\, *d_1}_{*d_1 \, \, 0})
({}^{\omega_1}_{\omega_2})>.
\label{eq:41}
\end{equation}
The reason for this specific choice of action $\tilde{S}(\omega_1,\omega_2)$ for
the doubled theory will become clear in the next section. An
obvious generalisation of the proceeding, with $T\to\tilde{T}=
({}^{\,\, 0 \,\,\, *d_1}_{*d_1 \, \, 0})$
shows that the partition function of the doubled theory
\[ \tilde{Z}(\lambda) = \int_{\Omega^1(M)\times\Omega^1(M)}{\cal D}
\omega_1{\cal D}\omega_2e^{-\lambda \tilde{S}(\omega_1,\omega_2)} \]
can be evaluated to obtain the square of (\ref{eq:40})
\begin{equation}
\tilde{Z}(\lambda)=\mid Z(\lambda) \mid^2 = (\frac{\lambda}{\pi})^{-1}
\label{eq:42}
\tau_{RS}(M).
\end{equation}
Note that there is no phase factor here. This is because the quantity
$\iota=d_++d_-$ for the action $\tilde{S}$ in (\ref{eq:41}) vanishes since
$\tilde{T}= ({}^{\,\, 0 \,\,\, *d_1}_{*d_1 \, \, 0})$
has a symmetric spectrum.

\section{The discrete version of the topological field theory}
\label{sec4}
We proceed to construct a discrete version of Abelian Chern-Simons
gauge theory.
The Whitney map enables the Abelian Chern-Simons theory to be discretised
by replacing the gauge field (1-form) $A\in\Omega^1(M)$ by the discrete
analogue, a 1-cochain $x\in C^1(K)$.

The most immediate way to do this is to construct the action $S_K$ of the
discrete theory by
\[ \lambda S_K(x) = \lambda S(W^K(x)) = \lambda\int_M d W^K(x)\wedge
W^K(x). \]
This can be shown to coincide with the discrete action for the Abelian
Chern-Simons theory introduced in \cite{albevario2}.
This prescription fails however, in the sense that the resulting
partition function $Z_K(\lambda)$ is not a topological invariant  i.e. is not 
independent of $K$, and does not reproduce the continuum expression for
the partition function. We demonstrate this by considering the resolvent
for $S_K$ obtained in an analogous way to the resolvent of the continuum
action $S$ described in the previous section. Let $T_K:C^1(K)\to
C^1(K)$ denote the self-adjoint operator on $C^1(K)$ determined by
\[ S_K(x)=\int_M dW^K(x)\wedge W^K(x) = < T_Kx,x>. \]
Then 
\[ {\rm Ker}(T_K) \subset {\rm Ker}(d_1^K). \]
Since for $x\in{\rm Ker}(d_1^K)$ we have
\begin{eqnarray}
<T_Kx,x> &=&\int_M dW^K(x)\wedge W^K(x), 
\label{eq:43} \\
&=& \int_M W^K(d_1^Kx)\wedge W^K(x). 
\label{eq:44}
\end{eqnarray}

Thus the discrete analogue of the resolvent (\ref{eq:31}) is a resolvent for $S_K$:
\[ 0\to{\mathbb{R}}\to^{\phi_0}\to\Omega^0(M)\to^{d_0^K}{\rm Ker}(d_1^K)
\subseteq{\rm Ker}(T_K)={\rm Ker}(S_K)\to 0. \]
The resulting partition function is the discrete analogue of the partition
function $Z(\lambda)$:
\[ Z_K(\lambda)={\det} '((\phi_0^K)^*\phi_0^K)^{-\frac{1}{2}}
{\det} '((d_0^K)^*d_0^K)^{\frac{1}{2}}
{\det} '(-\frac{i\lambda}{\pi}T_K)^{-\frac{1}{2}}.\]
In \cite{adhep} the following formula for $T_K$ was obtained:
\[ T_K [v_0, v_1] = \frac{1}{6}\sum[v_2,v_3]\]
where the sum is over all $1$-simplices $[v_2,v_3]$ such that 
$[v_0,v_1,v_2,v_3]$ is a $3$-simplex with orientation compatible with the
orientation of $M$.

It is possible to show \cite{adth} that $\det((\phi_0^K)^*\phi_0^K)=N_0^K={\rm dim }C^{(K)}=$
the number of vertices of $K$. Then the failure of the discretisation
prescription can be demonstrated by showing that the quantity
\[ \mid Z(\lambda) \mid^2 = \frac{1}{N_0^K}{\det} '(\partial_1^K
d_0^K){\det} '(T_K^2)^{-\frac{1}{2}} \]
is not independent of $K$. 

A discrete version of the doubled topological field theory with
action (\ref{eq:41}) has been constructed in \cite{adhep} in such a way that
the expression (\ref{eq:42}) for the continuum partition function is reproduced.
We briefly describe this in the following.

The discretisation prescription is
\begin{eqnarray}
(\omega_1,\omega_2)\in \Omega^1(M)\times\Omega^1(M)\to (x,y) \in C^1(K)
\times C^1(\hat{K}), 
\label{eq:45}
\\
S(\omega) = \lambda < ({}^{\omega_1}_{\omega_2}), ({}^{\,\, 0 \,\,\, 
*d_1}_{*d_1 \, \, 0})
({}^{\omega_1}_{\omega_2})> \to \tilde{S}_K(x,y)= \lambda <({}^x_y), 
({}^{\,\,\,\, 0 \,\,\,\,\,\,\,\, *^Kd^K}_{*^{\hat{K}}d^{\hat{K}}\,\,\, \, 0})({}^x_y)>, 
\label{eq:46}
\end{eqnarray}

where $K$ is the simplicial complex triangulating M, $\hat{K}$
is its dual, $C^q(K)$, $C^p(\hat{K})$, $d^K$ and $d^{\hat{K}}$ are
as described in the previous section. The analogue of the Hodge 
star operator is the duality operator $*^K$. This is
a map $*^K:C^q(K)\to C^{z-q}(\hat{K})$ ( and $*^{\hat{K}}:
C^p(\hat{K})\to C^{z-p}(K)$ which explains the need for field doubling
and the expression (\ref{eq:46}) for the discrete action $\tilde{S}_K(x,y)$. There is
a natural choice of resolvent for $\tilde{S}_K(x,y)$, analogous
to the resolvent (\ref{eq:31}) in the continuum case. It is 


The partition function is
\begin{equation}
\tilde{Z}_K(\lambda) = \int_{C^1(K)\times C^1(\hat{K})} {\cal D}x
{\cal D}y e^{-\tilde{S}_K(x.y)} .
\label{eq:47}
\end{equation}
Evaluating this by Schwarz's method with the resolvent above leads to:
\begin{eqnarray*}
\tilde{Z}_K(\lambda) =& (\frac{\lambda}{\pi})^{-1+N_0^K-N_1^K}\det((\phi_0^K)^*
\phi_0^K)^{-\frac{1}{2}}{\det}'((d_0^K)^*d_0^K)^{\frac{1}{2}}
{\det}'((d_1^K)^*d_1^K)^{-\frac{1}{4}}, \\
& \det((\phi_0^{\hat{K}})^* \phi_0^{\hat{K}})^{-\frac{1}{2}} 
{\det}'((d_0^{\hat{K}})^*d_0^{\hat{K}})^{\frac{1}{2}}{\det}'(
(d_1^{\hat{K}})^*d_1^{\hat{K}})^{-\frac{1}{4}}.
\end{eqnarray*}
There is no phase factor in (\ref{eq:47}) since $\zeta$ vanishes just like in (\ref{eq:42}). 
We have also used the fact that $\zeta=1-N_0^K+N_1^K$ which is shown in 
\cite{adhep}.

Now rewrite the determinant involving $\hat{K}$ - objects in terms
of determinants of $K$-objects. Modulo a possible sign $\pm$ we have
the formulae\cite{adhep}
\begin{eqnarray}
(*^{\hat{K}})^{-1} &=(*^{\hat{K}})^* = *^K, 
\label{eq:48} \\
(*^K)^{-1} &=(*^K)^* = *^{\hat{K}},
\label{eq:49} \\
(d_q^K)^* &=*^{\hat{K}}d^{\hat{K}}_{n-q-1}*^K,  
\label{eq:50} \\
(d_p^{\hat{K}})^* &= *^Kd_{n-p-1}^K*^{\hat{K}}. 
\label{eq:51}
\end{eqnarray}
( The $\pm$ signs are omitted because they will all cancel out in the 
following calculation.) Now:

\begin{eqnarray}
{\det}'((d_0^{\hat{K}})^*d_0^{\hat{K}}) 
&= {\det}'(*^Kd_2^K*^{\hat{K}} d_0^{\hat{K}}), \\
\label{eq:52}
&= {\det}'(*^{\hat{K}}(*^Kd_2^K*^{\hat{K}}d_0^{\hat{K}})*^K)  \\
&= {\det}'(d_2^K*^{\hat{K}}*^K),\\ 
\label{eq:53}
&= {\det}'(d_2^K(d_2^K)^*) \\ 
&= {\det} ((d_2^K)^*d_2^K), 
\label{eq:54}
\end{eqnarray}
and 
\begin{eqnarray}
{\det}'((d_1^{\hat{K}})^*d_1^{\hat{K}})  
&= {\det}'(*^Kd_1^K*^{\hat{K}} d_1^{\hat{K}}),  \\
\label{eq:55}
&= {\det}'(*^{\hat{K}}(*^Kd_1^K*^{\hat{K}}d_1^{\hat{K}})*^K) \\
&= {\det}'(d_1^K*^{\hat{K}}*^K), \\ 
\label{eq:56}
&= {\det}'(d_1^K(d_1^K)^*) \\
&= {\det} ((d_1^K)^*d_1^K). 
\label{eq:57}
\end{eqnarray}

The integration map (\ref{eq:37}) has a discrete analogue

\begin{eqnarray}
{\rm Ker }(d_2^K)^\perp &\to& {\mathbb{R}} , \nonumber \\
a &\to& <a,[M]>,
\label{eq:58}
\end{eqnarray}
where $[M]\in C_3(K)$, the orientation cycle of $M$, i.e. the sum of all
3-simplices of $K$, oriented so that their orientations are compatible
with the orientation of $M$. ( Note that $a\in{\rm Ker}(d_2^K)^\perp\subset 
C^3(K)$ can be evaluated on any element $\sigma\in C_3(K)$ to get a real
number $<a,\sigma>\in{\mathbb{R}}$.)
Define the map
\begin{equation}
\phi_3^K:{\mathbb{R}} \to{\rm Ker}(d_2^K)^\perp 
\label{eq:59}
\end{equation}
to be the inverse of (\ref{eq:58}). Then using the properties of $*^K$ and
$*^{\hat{K}}$, it can be shown that
\begin{equation}
\det ((\phi_3^K)^*\phi_3^K) = \det ((\phi_0^{\hat{K}})^*\phi_0^{\hat{K}})
^{-1}.
\label{eq:60}
\end{equation}
Now using (\ref{eq:60}), (\ref{eq:54}) and (\ref{eq:57}) we can rewrite (\ref{eq:47}) as
\begin{equation}
\tilde{Z}_K(\lambda) = (\frac{\lambda}{\pi})^{-1+N_0^K-N_1^K}
\tau_K(M),
\label{eq:61}
\end{equation}
where
\begin{equation}
\tau_K(M)=\det((\phi_0^K)^*\phi_0^K)^{-\frac{1}{2}}
\det((\phi_3^K)^*\phi_3^K)^{\frac{1}{2}}
\prod_{q=0}^2 {\det}'((d_q^K)^*d_q^K)^{-\frac{1}{2}(-1)^q} .
\label{eq:62}
\end{equation}
This quantity $\tau_K(M)$ is the R-torsion of the triangulation $K$ of $M$.
It is a combinatorial invariant of $M$ i.e.
is independent of the choice of triangulation $K$ \cite{cheeger,muller,dodzuik}.

This is the untwisted torsion of $M$, more generally the torsion can
be ``twisted'' by a representation of $\pi_1(M)$. The factors involving
the determinants $\det ' ((d_q^K)^*d_q^K)$ constitute the usual 
Reidemeister torsion of $M$ \cite{milnor}. When these are put together with the
factors involving $\det ((\phi_i^K)\phi_i^K)$, $i=0,3$, as in (\ref{eq:62})
we get the R-torsion ``as a function of the cohomology'' introduced and
shown to be triangulation-independent in \cite{RS}.

The expression (\ref{eq:62}) for $\tau_K(M)$ is analogous to the expression (\ref{eq:40})
for the R-torsion $\tau_{RS}(M)$, and in fact it has been
shown \cite{cheeger,muller} that these torsions are equal
\[ \tau_K(M)=\tau_{RS}(M). \]
It follows that the partition function (\ref{eq:61}) of the discrete theory coincides
with the partition function (\ref{eq:40}) of the continuum theory, except for the
$K$-dependent quantities $N_0^K$ and $N_1^K$ appearing in (\ref{eq:61}). These
quantities can be removed by a suitable $K$-dependent renormalisation
of the coupling parameter $\lambda$.

It is possible to show that \cite{adth}
\begin{eqnarray*}
\det((\phi_0^K)^*\phi_0^K)&=N_0^K\\
\det((\phi_3^K)^*\phi_3^K)&=\frac{1}{N_3^K}. 
\end{eqnarray*}
We will use this result in our numerical work.
\section{Numerical Results}
\label{sec5}
We are now in a position to proceed to numerically evaluate the discrete 
expressions for the torsion obtained.
This allows us to check the underlying theoretical ideas by numerically
verifying that the discrete expressions agree with expected analytic 
results. It also allows us to check that the results obtained are subdivision invariant.
The subdivision invariance of torsion is demonstrated by showing 
that if any
simplex of the triangulation is subdivided, the value of the torsion does not 
change. This is what is meant by topological invariance in the
discrete setting. The expected analytic result for torsion for a lens space
$L(p,1)$ is (See \cite{RS}) 
\[ T(L(p,1))=\frac{1}{p}. \]
Thus $T(S^3)=T(L(1,1))=1$.

We also show, numerically, that the discrete expression
for the Chern-Simons partition function obtained without using the 
* operator is not a topological invariant.
This shows very clearly the importance 
of the doubling construction method used
in the discretisation method, for capturing topological information.

In order to proceed, we need to efficiently 
triangulate the spaces $S^3$ and L(p,q). First we triangulate $S^3$. 
We do this by considering a four dimensional simplex $[v_0,v_1,v_2,v_3,v_4]$ 
and observing that the boundary of this object is precisely the triangulation
$K$ of $S^3$ which we require. Next we turn to spaces L(p,q) which we need to triangulate
in order to proceed. An efficient
triangulation of this space has been constructed by Brehm and Swiatkowski
\cite{BS}. We use this procedure for our computations \cite{samik}.

We can now summarise our numerical results.
The R-torsion for a simplicial complex $K$, with dual cell
complex $\hat{K}$, for either $S^3$ or L(p,1) involves
evaluating
\[ T=\sqrt{\frac{1}{N_0 N_3}\det(\partial_1 d_0)\det(\partial_2d_1)^{-1}
\det(\partial_3d_2)}, \]
where $N_i=$nos.\ of $i$-simplices in $K$, $\partial_1, \partial_2, \partial_3$ 
are boundary operators on $K$ and $d_0,d_1,d_2$ are coboundary operators on $K$.
Note that, in the discrete setting, these operators can be expressed as matrices.
We do this by using the vertices, edges, faces
and tetrahedra of our complex as a basis list. Any $p$-simplex in the complex can
then be expressed in terms of this.
Since we know what $d$ maps the various basis list
elements to,  we can set the coefficients of its matrix representation.
When we say $\det(\partial_1 d_0)$, for example,
we simply mean the determinant of the matrix which results from multiplying the matrices 
corresponding to the operators $\partial_1$ and $d_0$.

If our complex consisted of just one triangle $[0,1,2]$, the basis list
would be:

\[
\begin{array}{lc}
0 & [0], \\ 
1 & [1], \\
2 & [2], \\
3 & [0,1], \\
4 & [0,2], \\
5 & [1,2], \\
6 & [0,1,2].
\end{array}
\]

We know that $d[0]=[1,0]+[2,0]$. This can be expressed in terms of our basis list  
as $d$ acting on basis element $0$ going to $-3-4$. So the matrix $d$ for this
complex is:

\[
\left( \begin{array}{ccccccc}
0 & 0 & 0 & 0 & 0 & 0 & 0 \\
0 & 0 & 0 & 0 & 0 & 0 & 0 \\
0 & 0 & 0 & 0 & 0 & 0 & 0 \\
-1 & 1 & 0 & 0 & 0 & 0 & 0 \\
-1 & 0 & 1 & 0 & 0 & 0 & 0 \\
0 & -1 & 1 & 0 & 0 & 0 & 0 \\
0 & 0 & 0 & 1 & -1 & 1 & 0 \\
\end{array} \right). \]

Note that the last column is zero since $d$ has nothing to map a $2$-simplex to and
that the first three rows are zero since nothing is mapped to $0$-simplices.
If we act on $0$ with this matrix we get $-3-4$, as expected, and $d^2=0$.

If we do not use $\hat{K}$ then from section~\ref{sec4},
\[ \hat{T} = \mid Z(1)\mid^2=
\frac{1}{N_0^K}({\det}' \partial_1^K d_0^K)
(\det T_K^2)^{-\frac{1}{2}}. \]

We evaluated the quantity $\hat{T}$ numerically for various triangulations $K$
of $S^3$ and found the following results shown in Table~\ref{tab2} for the change
of $\hat{T}$ under subdivision where $m-n$
corresponds to a triangulation of $S^3$ with $n$ vertices and $m$ tetrahedra
and where $X_i= \det \partial_i d_{i-1}$.
\\ 

\begin{table}
\begin{center}
\begin{tabular}{|c|c|c|c|c|c|c|} \hline
Complex & $X_1$ & NEW &
$\frac{X_1*X1}{NEW} $ & $N_3$ & $N_0$ & $Z^4$\\
\hline
5-5 &   125&    15625      & 1 & 5     & 5 & 0.04 \\
\hline
8-6&    1152    &2985984   & 0.444  &8    & 6 & 0.0123 \\
\hline
9-6 &   2304 &  15116540  & 0.351  & 9    & 6 & 0.00975\\
\hline
\end{tabular}
\end{center}
\caption{Results for $T$.}
\label{tab2}
\end{table}

It is clear that $\hat{T}$ for $S^3$ is not subdivision invariant. 
We next evaluate $T$ for $S^3$ and for L(p,1) and check that it is indeed 
subdivision invariant and agrees with the analytic calculations for
L(p,1), with p=2,3,4 and 5. These results are shown in Tables~\ref{tab4}-\ref{tab6}.

As a check on the numerical method we
also count the number of zero modes of the Laplacian operator
on the different p-chain spaces. These numbers gives the dimension
of the Homology groups and are shown in Table~\ref{tab3}.

\begin{table}
\begin{center}
\begin{tabular}{|c|c|c|}\hline
p & No. of zero-modes of $\Delta_p$ & Dim $H^p$ \\
\hline
0 & 1 & 1 \\
1 & 0 & 0 \\
2 & 0 & 0 \\
3 & 1 & 1 \\
\hline
\end{tabular}
\end{center}
\caption{Zero modes for $S^3$.}
\label{tab3}
\end{table}

\subsection{$S^3$}

As a further check, a systematic way of carrying out subdivision known as
the Alexander moves \cite{Alex} was used to study the subdivision invariance of the torsion.
There are four such moves in 3 dimensions. They are best explained by 
example. We have
\begin{enumerate}
\item $[0,1,2,3,4]->[x,1,2,3,4] + [0,x,2,3,4]$. 
\item $[0,1,2,3,4]->[x,1,2,3,4] + [0,x,2,3,4] + [0,1,x,3,4]$. 
\item $[0,1,2,3,4]->[x,1,2,3,4] + [0,x,2,3,4] + [0,1,x,3,4] + [0,1,2,x,4] $. 
\item $[0,1,2,3,4]->[x,1,2,3,4] + [0,x,2,3,4] + [0,1,x,3,4] + [0,1,2,x,4] + [0,1,2,3,x]$.
\end{enumerate}
These are all natural operations in that the first (nth) move corresponds
to adding a vertex splitting the 1-simplex (n-simplex) $[0,1]$ ($[0,1,\dots,n]$)
and connecting it to all the vertices resulting in two (n+1) tetrahedra. 

The torsion T, in terms of its component determinants,
and the way they change under the Alexander moves is
exhibited in the table below,

\begin{table}
\begin{center}
\begin{tabular}{|c|c|c|c|c|c|c|c|} \hline  
Complex & $X_1$ & $X_2$ & $X_3$ &
$\frac{X_1*X_3}{X_2} $ 
& $N_3$ & $N_0$ & T\\ 
\hline  
s3 & 625 & 15625 & 625 & 25 & 5 & 5 & 1 \\ 
a1s3 & 5184 & $2.985984*10^6$ & $2.7648*10^4$ & 48 & 8 & 6 & 1 \\
a2s3 & 7776 & $1.511654*10^7$ & $1.04976*10^5$ & 54 & 9 & 6 & 1\\
a3s3 & 5184 & $ 2.985984*10^6$ & $2.7648*10^4$ & 48 & 8 & 6 & 1\\
a4s3 & 625 & 15625 & 625 & 25 & 5 & 5 & 1 \\
\hline 
\end{tabular}
\end{center} 
\caption{R-torsion for $S^3$.}
\label{tab4}
\end{table}

\noindent
where $X_i=\det \partial_i d_{i-1}$ 
and $a2s3$ means the complex which resulted after the type 2 Alexander
moves were performed on the $S^3$. It's clear from this that
\[ T=\frac{1}{N_3 N_0}\det(\partial_1 d_0) ( \det \partial_2 d_1)^{-1} \det
(\partial_3 d_2) \] 
is subdivision invariant and thus a topological invariant of the manifold.

As a final check we tried several other subdivisions.
We took a given triangulation and barycentric subdivided one or more ($n$)
of its faces to get $1s3, 2s3, \dots$ ($ns3$). 
The results are in Tables~\ref{tab4} and \ref{tab5}.

\begin{table}[h]
\begin{center}
\begin{tabular}{|c|c|c|c|c|c|c|c|} \hline  
Complex & $X_1$ & $X_2$ & $X_3$ &
$\frac{X_1*X_3}{X_2} $ & $N_3$ & $N_0$ & T\\ 
\hline  
s3 & 625 & $1.562*10^4$ & 625 & 25 & 5 & 5 & 1 \\ 
1s3 & 5184 & $2.985984*10^6$ & $2.7648*10^4$ & 48 & 8 & 6  & 1\\
2s3 & $3.9448*10^4$ & $5.2166*10^8$ & $1.018325*10^6$ & 77 & 11 & 7 & 1 \\
3s3 & $2.79936*10^5$ & $8.418024*10^{10}$ & $3.36798*10^7$ & 112 & 14 & 8 & 1 \\
4s3 & $1.876833*10^6$ & $1.266709*10^{13}$ & $1.032626*10^9$ & 153 & 17 & 9 & 1 \\
5s3 & $1.270325*10^7$ & $2.037398*10^{15}$ & $3.207653*10^{10}$ & 200 &20 & 10 & 1 \\
\hline 
\end{tabular}
\end{center} 
\caption{R-torsion for $S^3$.}
\label{tab5}
\end{table}


\subsection{Lens spaces}
We conclude with the results for the lens spaces. The results are
shown in Table~\ref{tab6} .  As can be seen, these agree extremely well with the known analytic
result T(L(p,1))=$1/p$.

\begin{table}
\begin{center}
\begin{tabular}{|c|c|c|c|c|c|c|c|} \hline  
Complex & $X_1$ & $X_2$ & $X_3$ &
$\frac{X_1*X_3}{X_2}$ & $N_3$ & $N_0$ & T\\ 
\hline  
L(2,1) & $1.062937*10^{10}$ & $3.618662*10^{29}$ & $3.74484*10^{21}$ &
110 & 40 & 11 & 1/2 \\
L(3,1) & $9.108*10^{12}$ & $1.143589*10^{43}$ & $1.08817*10^{32}$ &
86.666 & 60 & 13 & 1/3 \\
L(4,1) & $1.1027*10^{16}$ & $4.0468*10^{58}$ & $2.89*10^{44}$ &
78.7487 & 84 & 15 & 1/4 \\
L(5,1) & $1.790617*10^{19}$ & $1.761432*10^{76}$ & $7.49187*10^{58}$ &
76.16 & 112 & 17 & 1/5 \\
\hline
\end{tabular}
\end{center}
\caption{R-torsion for lens spaces.}
\label{tab6}
\end{table}

\section{Conclusions}
\label{sec6}
The method of discretisation introduced works extremely well.
The main point of the method is to construct discrete analogues for the set 
$(\Omega^p,d,\wedge,*)$. Previous work in the direction has 
neglected the Hodge star operator, * \cite{albeverio,BR}. We have thus 
demonstrated that the Hodge star operator plays a 
vital role in the construction of topological
invariant objects from field theory. We were able to construct 
an expression for the partition function which is correct even
as far as overall normalisation is concerned. Mathematically the 
equivalence between the Ray-Singer torsion and the combinatorial 
torsion of Reidemeister was proved independently in 1976
by Cheeger \cite{cheeger} and M\"{u}ller \cite{muller}. It is nice to see the 
result emerge
in a direct manner by a formal process of discretisation. On the way
we also had to double the original system so that K, the triangulation
, and $\hat{K}$, its dual, are both present. If this doubling and
the reason for it are overlooked, then the topological 
information present in the discretisation is lost, as our numerical
results demonstrated.

It is clear that the geometry motivated discretisation method
introduced is very general
and that it can be used to analyse a wide variety of physical systems.
In the approach outlined we have captured topological features. In applications
capturing geometrical features of a problem are also very important. We are 
currently investigating this aspect of our approach.

A limitation of the method is that there is no simple generalisation to deal 
with non-abelian theories. The discrete analogies of $d$, $d^\dagger$ were linear :
there is no natural discrete analogue of $d_A:=d+A$, with $A$ a Lie algebra valued
one form.

\section*{Acknowledgments}
The work of S.S. is part of a project supported by Enterprise Ireland. 
D.A. would like to acknowledge support from Enterprise Ireland and Hitachi. 
Samik Sen and S.S.
would like to thank D. Birmingham for explaining the triangulation method 
of Ref. \cite{BS}.

\end{document}